\documentclass[conference]{IEEEtran} 

\usepackage{fancyhdr, epsfig, epsf, amsthm, amsmath, amssymb, amsfonts, subfigure, color}
\usepackage{dblfloatfix} 
\usepackage{threeparttable}
\usepackage[noadjust]{cite}
\usepackage{dsfont}
\usepackage{enumerate}
\usepackage{comment}

\newtheorem{lemma}{Lemma}

\newtheorem{proposition}{Proposition}

\def\qed{$\blacksquare$}

\def\endproof{\hfill \qed}

\def\SINR{\mathsf{SINR}}

\def\l{\left}
\def\r{\right}
\def\({\left(}
\def\){\right)}

\def\[{\left[}
\def\]{\right]}

\def\EE{\mathsf{EE}}

\def\papertitle{User-Centric Mobility Management in Ultra-Dense Cellular Networks under Spatio-Temporal Dynamics}
\IEEEoverridecommandlockouts

\begin{document}

\title{ \fontsize{19.6}{21}\selectfont \papertitle}



\author{Jihong Park, Sang Yeob Jung, Seong-Lyun Kim, Mehdi Bennis$^{\dagger}$, and M\'erouane Debbah$^{\circ}$
\thanks{J. Park, S. Jung, and S.-L. Kim are with School of Electrical \& Electronic Engineering, Yonsei University, Seoul, Korea (email: \{jhpark.james, syjung, slkim\}@ramo.yonsei.ac.kr). M. Bennis$^{\dagger}$ is with Centre for Wireless Communications, University of Oulu, Finland (email: bennis@ee.oulu.fi). M. Debbah$^{\circ}$ is with Mathematical and Algorithmic Sciences Lab, France Research Center, Huawei Technologies Co. Ltd. (email: merouane.debbah@huawei.fr).}}

\maketitle \thispagestyle{empty}

\begin{abstract} This article investigates the mobility management of an ultra dense cellular network (UDN) from an energy-efficiency (EE) point of view. Many dormant base stations (BSs) in a UDN do not transmit signals, and thus a received power based handover (HO) approach as in traditional cellular networks is hardly applicable. In addition, the limited front/backhaul capacity compared to a huge number of BSs makes it difficult to implement a centralized HO and power control. For these reasons, a novel user-centric association rule is proposed, which jointly optimizes HO and power control for maximizing EE. The proposed mobility management is able to cope not only with the spatial randomness of user movement but also with temporally correlated wireless channels. The proposed approach is implemented over a HO time window and tractable power control policy by exploiting mean-field game (MFG) and stochastic geometry (SG). Compared to a baseline with a fixed HO interval and transmit power, the proposed approach achieves the $1.2$ times higher long-term average EE at a typical active BS.
\end{abstract}


\section{Motivation and Contributions}
Imagine energy-saving ceiling lights adaptively turn on when walking down a dark corridor. This resembles a cellular network operation where a huge number of base stations (BSs) are deployed \cite{TowardUDN:15,JHParkTWC:15, JHParkPIMRC:16}. To maximize energy-efficiency (EE), many user void BSs stand by in \emph{dormant} mode to be \emph{active} as soon as a mobile user arrives within their coverage \cite{NCT2013}. 

It is then natural to ask how to make energy-efficient cell coverage in such a ultra-dense cellular network (UDN), by adjusting a handover (HO) rule and power control policy. In addition, implementing energy-efficient HO and power control in a distributed manner is its prime task. A centralized control as in traditional cellular networks is hardly applicable due to the limited front/backhaul capacity compared to the number of BSs in a UDN. 

To resolve these issues, this article proposes a user-centric association rule (see Fig. 1) with a decentralized power control policy for maximizing BS's EE. A notable point is the proposed scheme incorporates user movement and temporal wireless channel dynamics under a large number of BSs. Regardless of this complicated nature, its operation is strikingly simple and tractable (see Fig. 2) by the aid of a spatio-temporal network dynamics framework, proposed in our preceding work \cite{UdnMFGSG:GC16} on the basis of mean-field game (MFG) and stochastic geometry (SG).

\begin{figure}\centering
\includegraphics[width=7cm]{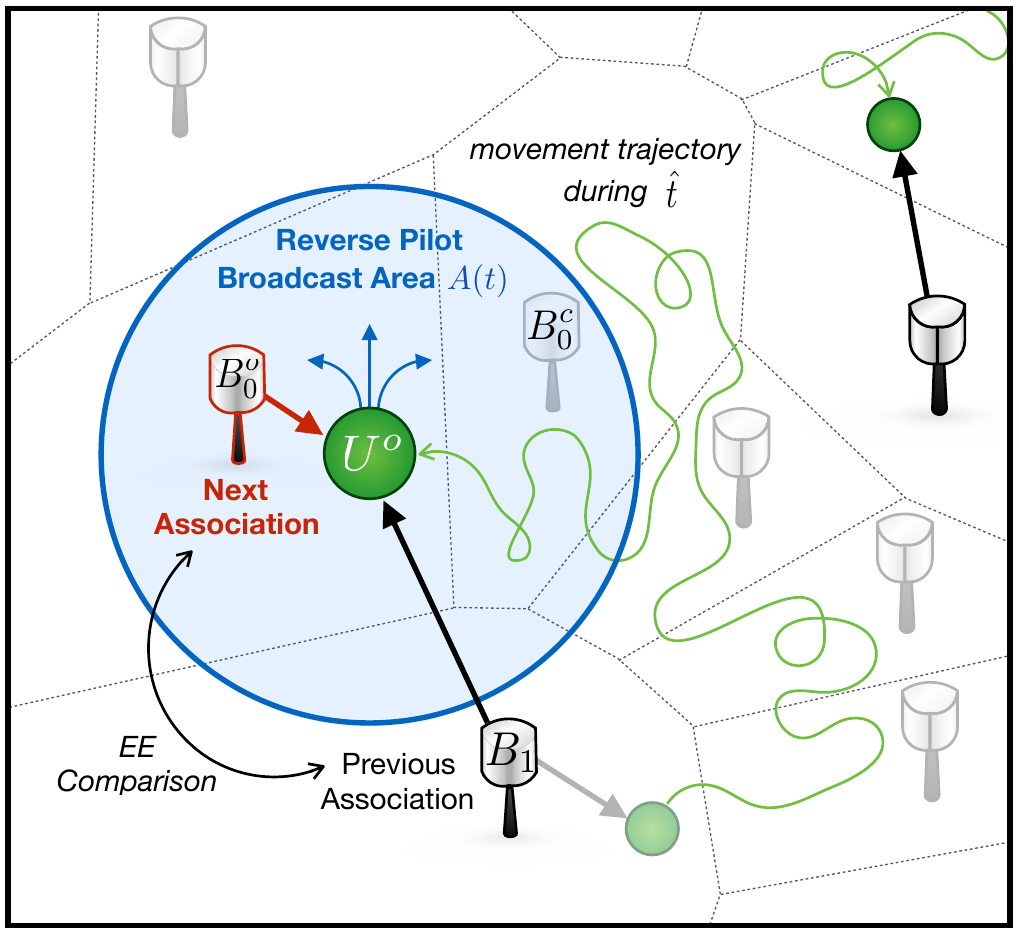}
	\caption{User-centric reverse association illustration. At HO intervals of $\hat{t}$, a typical user $U^o$ broadcasts a reverse pilot to dormant BSs. Each dormant BS within $A(\theta,t)$ receives the pilot, and responds back to $U^o$ if its associating with $U^o$ increases EE. $U^o$ associates to the nearest $B_0^o$ out of the responded BSs, and disconnects to the previous association with $B_1$.}
\end{figure}

The contributions of this paper are as follows.
\begin{itemize}
\item User-centric reverse association in a UDN is proposed.
\item Closed-form EEs are derived under Brownian motion (BM) mobility with temporally correlated fading (Propositions 2 and Corollary 1).
\item EE optimal power control policy and HO rule are proposed in tractable forms (Propositions 3 and 4).
\item Mean-field (MF) interference convergence is proved in a UDN under BM mobility (Proposition 1).
\end{itemize}




\begin{figure*}
	\centering
	\subfigure[Without utilizing MF interfrence]{\includegraphics[width=7cm]{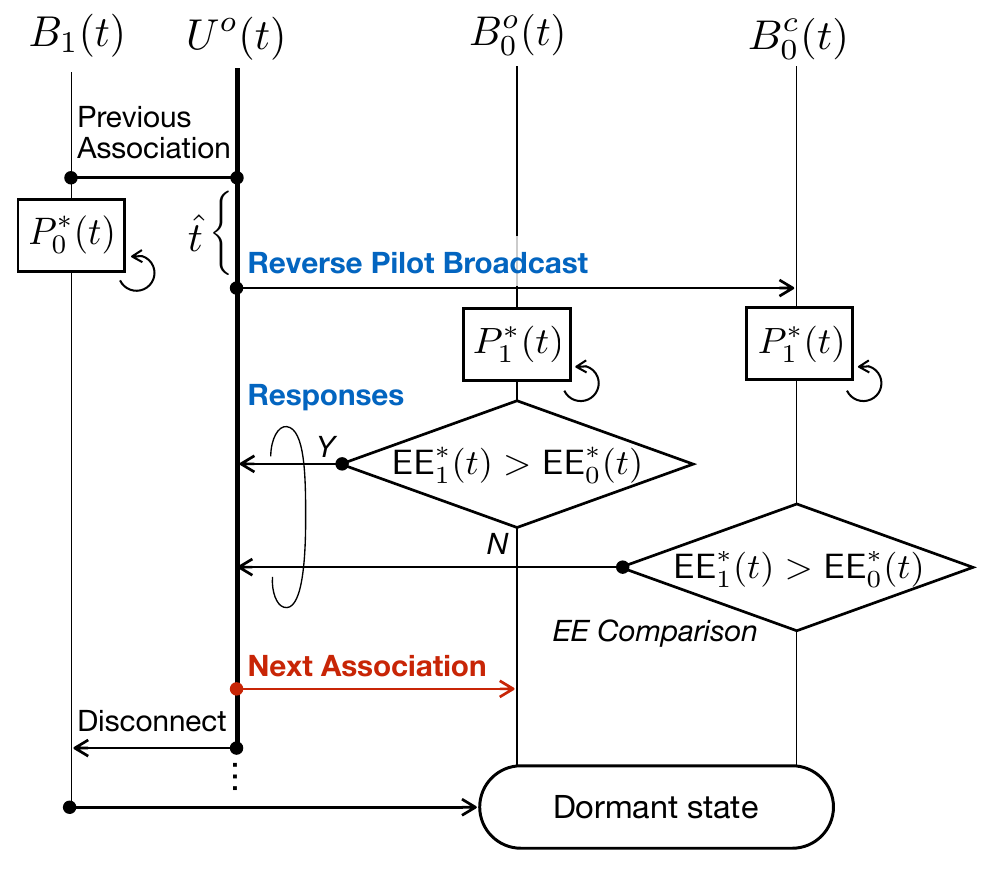}} 
	\subfigure[With utilizing MF interference]{\includegraphics[width=7cm]{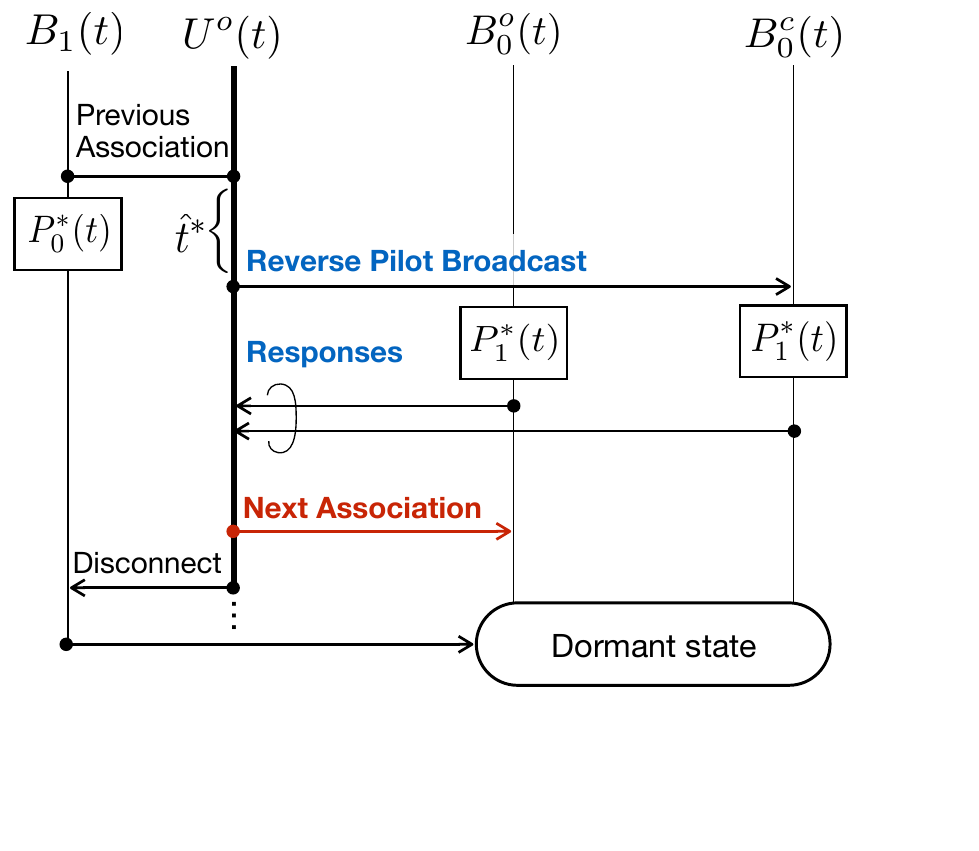}}
	\caption{User-centric reverse association flow charts with and without utilizing MF interference 
where $U^o$ is a typical user, $B_1(t)$ is $U^o$'s associated BS, $B_0^o(t)$ $U^o$'s nearest BS at time $t$, and $B_0^c(t)$ an arbitrary BS except for $B_1(t)$ and $B_0^o(t)$ within the pilot broadcast area $A(t)$ (see Section III-B. for details)}.
\end{figure*}

\section{System Model} \label{Sect:SysModel}
\textbf{User Mobility and Network Deployment}.
Initial user locations $y(t)=\{y_x(t),y_y(t) \}$ at time $t=0$ follow a homogeneous Poisson point process (PPP) with density $\lambda_b$. At $t>0$, the users move according to a Brownian motion (BM) represented by a stochastic differential equation (SDE)
\begin{align}
dy(t) &= \frac{v^2}{\pi} d\mathbb{W}(t) \label{Eq:Mobility}
\end{align}
where the standard Wiener process $\mathbb{W}(t)= \{W_x(t), W_y(t)\}$ for mutually independent $W_x(t), W_y(t)\sim \mathcal{N}(0, t)$ and $v$ is the mean displacement distance per unit time, interpreted as average velocity. BS locations follow a homogeneous PPP with density $\lambda_b$. BSs with no user in their coverages become dormant.

\textbf{Channel Dynamics}. Transmitted signals from active BSs experience path-loss attenuation accompanied by spatially i.i.d. but temporally correlated fading. The fading coefficient vector at an arbitrary user is given as $g(t)={\{g_x(t), g_y(t)\}}$. The fading coefficients are constants within a block, and evolves from block to block while satisfying the Markov property, i.e. Gauss-Markov correlated block fading \cite{Honig:12}. Consider the block length is infinitesimal, leading to the following fading evolution law \cite{Meriaux:13}:
\begin{align}
d g(t)&= \frac{1}{2}\[\mu - h(t)\] dt + \eta \cdot d  \mathbb{W}(t) \label{Eq:Channel}
\end{align}
where $\mu:=\{\mu_x, \mu_y\}$ for non-negative constants $\mu_x, \mu_y$, $0\leq \eta < \infty$, and $\mathbb{W}_i(t)$'s are mutually independent Wiener processes. Adjusting $\mu$ and $\eta$ is able to capture different channel statistics including Rayleigh fading for $\mu=h(t)$ and Rician fading for $t\rightarrow \infty$ \cite{Meriaux:13, UdnMFGSG:GC16}. The path-loss attenuation for distance $x$ follows $\min\(1, x^{-\alpha}\)$ for path-loss exponent $\alpha>2$. Each user receives signals from active BSs located within a reception distance $R$ \cite{UdnMFGSG:GC16}, providing the received powers larger than noise floor $\sigma^2$.

\textbf{Power Consumption}. Dormant BSs consume $P_c$ Joule (J) per unit time for powering circuits to receive user requests. Each active BS additionally consumes $P$ for data transmissions. The value $0<P\leq P_{\max}$ is distributedly controlled based on each active BS's associating and interfering channel states so as to maximize its own utility in the mean-field regime to be characterized in Section IV-B. Different active BSs transmit the same power when their channel states are respectively identical, i.e. homogeneous admissible controls for transmit powers \cite{Meriaux:13}.

\section{User-Centric Reverse Association for Energy-Efficient Ultra-Dense Cellular Networks}
In this article we focus on investigating the dormant BS impact on HO and power control policies in a UDN. For this purpose, we hereafter consider that $\lambda_b/\lambda_u\gg 1$ such that the following approximations hold: (i) an arbitrary BS's active probability $p_a$ is $\lambda_u/\lambda_b$ and (ii) a user's association distance is his displacement distance. These approximation become exact calculations when $\lambda_b/\lambda_u \rightarrow \infty$ \cite{JHParkTWC:15}.

Notwithstanding some compromises on practical feasibility, the above approximations provide the logarithmically increasing average rate tendency as densification increases \cite{JHParkTWC:15}, which was also proven in realistic scenarios by simulation \cite{TowardUDN:15}. In addition, the average rate under the assumptions can directly be regarded as an approximation for a BS density by adjusting a constant that is only a function of path-loss exponent \cite{JHParkPIMRC:16}. In such a UDN, the following subsections define EE and propose a user-centric association.

\subsection{Energy-efficient HO and power control}
As the utility function for HO decision, consider each BS's EE, downloaded bits per energy consumption in unit of bits/J. Precisely, define EE as the downloaded data volume per unit time at a target receiver, divided by the power consumption of the transmitter associated with the target receiver.

When a user tries to associate with a dormant BS, from the dormant BS point of view, consider $\EE_1(t)$ as the EE when the dormant BS allows the user association and $\EE_0(t)$ as the current EE while maintaining its dormant operation. By the above EE definition, $\EE_1(t)$ and $\EE_0(t)$ are represented as

\vspace{-10pt}\small\begin{align}
\EE_0(t) &= \frac{\gamma_0\(t, P_0(t)\)}{P_c} \\
\EE_1(t) &= \frac{\gamma_1\(t, P_1(t)\)}{P_c + P_1(t)}
\end{align}\normalsize
where $P_c>0$ is the constant circuit power consumption, $\gamma_0\(t, P_0(t)\)$ is the average rate given by the user's currently associated BS with transmit power $P_0(t)$, and $\gamma_1\(t, P_1(t)\)$ is the predicted average rate if the target dormant BS turns on and associates with the user with transmit power $P_1(t)$.

A BS's HO decision is carried out by comparing $\EE_0(t)$ and $\EE_1(t)$. If the BS accepts the HO, the average rate increases by reducing the association distance in return for additional $P_1(t)$ power consumption. To resolve this trade-off, each dormant BS accepts HO when its $\EE_1(t)>\EE_0(t)$.

Now that this EE comparison depends on transmit powers, the dormant BS predicts its optimal transmit power $P_1^*(t)$ that maximizes $\EE_1(t)$. Active BSs also transmit with their own EE maximizing powers. Let $\EE_a(t)$ denote such a typical active BS's EE, given as
\begin{align}
\EE_a(t) = \frac{\gamma_0(t,P_0(t))}{P_c + P_0(t)} = \EE_0(t) \(1 + \frac{P_0(t)}{P_c}\)^{-1} .
\end{align}

The optimal transmit power $P_0^*(t)$ maximizing $\EE_a(t)$ is used for calculating the $\EE_0(t)$'s optimum value at the EE comparison. The proposed HO decision and power control policies are integrated in the next subsection via a user-centric association.

\begin{figure*}[b]
\hrulefill
\setcounter{equation}{6}
\small\begin{align}
\EE_1(t) &=   \frac{\max\(1-\varepsilon \exp\l\{-\lambda_b A(\theta,t) \r\}, 0\) }{P_c + P_1(t) } \[c_1   + \log\(\frac{P_1(t) }{\sigma^2 N^{-1} {  \[\lambda_b \exp\(-\lambda_b A(\theta,t){N_p}^{-1}\) \]}^{-\frac{\alpha}{2}} + \hat{I}_1(t)} \)\] \label{Eq:EE1} \\
\EE_0(t) &= \frac{1}{P_c}  \[  c_1 +    \log\(\frac{P_0(t) }{\(2 v\sqrt{ \hat{t} }\)^\alpha\[\sigma^2 N^{-1} + \hat{I}_0(t)\]} \) \]  \label{Eq:EE0}
\end{align} 
where
\vspace{-10pt}\begin{align}
\hat{I}_i(t) &=   \frac{(\lambda_u \pi R)^2 \hat{P}_i(t)}{\sqrt{N}{\lambda_b}^\frac{\alpha}{2}} \(1 + \frac{1-R^{2-\alpha}}{\alpha-2}\)   \[ 2 \eta^2\(1-e^{-t}\)^2 + |\mu|^2\(1-e^{-\frac{t}{2}}\)^2 \]    \quad \text{for $i\in\{0, 1\}$}, 
\label{Eq:I_i}\\
A(\theta,t) &=\pi\( \frac{1}{\theta}\[ 2 \eta^2\(1-e^{-t}\)^2 + |\mu|^2\(1-e^{-\frac{t}{2}}\)^2 \] \)^{\frac{2}{\alpha}}, \label{Eq:AreaPilot}\\
c_1 &=\frac{\alpha \(\gamma + \log\pi\)}{2} + 2 \int_0^\infty Q_1\(\frac{|\mu|}{\eta\(1 + e^{-\frac{t}{2}}\)}, \frac{e^x}{\eta\(1-e^{-t}\)} \)dx , 
\end{align}
\small $\hat{t}$ is the elapsed time after HO (i.e. $t - \hat{t}\lfloor t/\hat{t}\rfloor$), $\sigma^2$ is the noise power, $\gamma\approx 0.5771$ is the Euler constant, $0 \leq P_i(t), \hat{P}_i(t) \leq P_{\max}$, $\varepsilon>0$, $Q_1(a, b):=\int_b^\infty x e^{-\frac{x^2 + a^2}{2}}I_0(ax)dx$ is a Marcum Q-function, and $I_0(x)$ is the first kind modified Bessel function of order zero \cite{AbramowitzBook:HandbookFunctions:1995}. \normalsize
\end{figure*}

\subsection{User-centric reverse association} \label{Sect:Association}
Consider the following procedure of an EE-optimal user association initiated by users (see Fig. 2-a), instead of the BSs as in traditional cellular networks.

\begin{itemize}
\item[1.] \textbf{Reverse Pilot Broadcast}. At time $t$, a user broadcasts a pilot signal to dormant BSs at intervals of $\hat{t}>0$.
\item[2.] \textbf{Response after EE Comparison}. Each dormant BS that received the pilot compares its $\EE_0^*(t)$ as previous EE with the optimized transmit power $P_0^*(t)$ and its $\EE_1^*(t)$ as the predicted EE when allowing the user association with the optimal power $P_1^*(t)$. If $\EE_1^*(t)> \EE_0^*(t)$, then it responds back to the user; otherwise, it remains dormant.
\item[3.] \textbf{Next Association}. The user makes an association with the nearest responding BSs by comparing the BS response signal powers. If no response signal exists, the user maintains his previous association until sending the next reverse pilot.
\end{itemize}

The following addresses major issues in the proposed association.


\textbf{Power Control and EE Comparison Complexity}.
Calculating $P_1^*(t)$ is solved by a stochastic differential game (SDG) via Hamilton-Jacobi-Bellman partial differential equation (HJB) \cite{Meriaux:13}. The problem is that each BS's power control affects the other BSs' through interference. For this reason $P_1^*(t)$ calculation requires to solve a set of HJBs. MFG at this point plays a key role in approximating the problem, as long as sufficient BS-to-user density ratio and/or number of antennas are ensured, MFG allows BSs to immediately calculate $P_1^*(t)$ by only considering a single power control in MF interference, instead of considering individual BS's power control. Such a MF interference approach is also applicable to calculating $P_0^*(t)$ for $\EE_0^*(t)$. This significantly reduces the power control complexity. Furthermore, MF interference enables us to derive closed-form $\EE_0(t)$ and $\EE_1(t)$, thereby replacing the EE comparison procedure at the second step by simply calculating an EE optimal HO interval, denoted as HO time window $\hat{t}^*$ (see Fig. 2-b). 

\textbf{Wasteful HO Trials}. The HO time window for individual BS's EE may bring about too frequent HO trials. This results in the overlapping HO requests consecutively to the same BS and/or ping-pong HO, degrading the overall network performance. To avoid it, the minimum HO time window is prepared (see Proposition 4), defined as the HO interval so that a typical user's movement distance is larger than his nearest BS distance with probability $\beta$. This concept is in line with a time-to-trigger (TTT) method in traditional cellular netwroks.

\textbf{Pilot Response Congestion}. Pilot response receptions at the second step may encounter control signal congestion. The reason comes not only from the limited number of pilot response resource blocks $N_p$ but also from decentralized responding operation under a large number of dormant BSs, which is similar to random access congestion in traditional cellular networks \cite{HLee:13}. This can be mitigated by reducing the pilot signal transmit power at the first step while assigning a certain reference signal received power (RSRP) threshold $\theta$ at the second step so that only the RSRP threshold exceeding dormant BSs can respond. This paper captures its impact on EE (see Proposition 2 and Fig. 4), but defers the optimal design of $\theta$ to future work.

\section{Mobility Management under Mean-Field Interference in Ultra-Dense Cellular Networks}

This section aims at providing an energy-efficient mobility management based on the user-centric association in a UDN. 
\subsection{Preliminaries}
A typical user's mobility is described by using its movement distance distribution as follows.
\begin{lemma} \emph{(BM Movement Distance)} Under the BM mobility in \eqref{Eq:Mobility}, the complementary cumulative distribution function (CCDF) of an arbitrary user's movement distance during $t>0$ is
\vspace{-10pt}\setcounter{equation}{5}\small\begin{align}
\Pr\(U(t)> u\) &= \exp\(- \frac{\pi u^2}{4 t v^2}  \r).
\end{align}
\end{lemma}\normalsize
\noindent\emph{Proof:}
Let $U(t)$ decompose into $u_x(t)$ and $u_y(t)$ such that $U(t) = \sqrt{u_x(t)^2 + u_y(t)^2}$. Solving a Fokker-Planck-Kolmogrov (FPK) equation \cite{ReveszBook:RW} satisfying \eqref{Eq:Mobility} leads to the distributions of $u_x(t)$ and $u_y(t)$ independently and identically following $\mathcal{N}(0, 2 v^2 t/\pi)$. By the relationship among $U(t)$, $u_x(t)$, and $u_y(t)$, $U(t)\sim$ Rayleigh$\(0, v\sqrt{2t/\pi}\)$, and applying its CCDF completes the proof. \endproof \vspace{5pt}

This movement distance can also be represented by using a typical user's nearest BS distance $D(t)$. Comparing Lemma 1 with the CCDF of $D(t)$, $\Pr\(D(t)>d\) = \exp\(-\lambda_b \pi d^2\)$ \cite{HaenggiSG}, leads to the following relationship between $U(t)$ and $D(t)$.

\begin{lemma} \emph{(Movement-to-Nearest Distance Conversion)} BM movement distance during $t$, $U(t)$, is identical to a typical user's nearest BS distance $D(t)$ under BS density $1/\(4 t v^2\)$.
\end{lemma}

From a network point of view, user locations under BM mobility preserves their uniform distribution as specified below.
\begin{lemma} \emph{(Spatial Homogeneity)} If each user moves independently of the current location and their locations at $t=0$ follow a homogeneous PPP with density $\lambda_u$, then for all $t>0$ user locations follow another homogeneous PPP with $\lambda_u$.
\end{lemma}
The above result is stated in \cite{Haenggi:GC10}, and can be directly derived by using the displacement theorem \cite{Kingman:Book} or
Theorem 1.1 in \cite{ReveszBook:RW}. This lemma allows us to utilize traditional stochastic geometric approaches for both associating and interfering links from the perspective of a typical user under BM mobility.

\begin{figure*}[b]
\hrulefill
\setcounter{equation}{11}
\small\begin{align}
P_1^*(t) &= \min\( \frac{P_{c}}{W\left(P_{c}\exp\l\{ c_{1}-1 -\log\(   \sigma^2 N^{-1} {  \[\lambda_b \exp\(-\lambda_b A(\theta,t){N_p}^{-1}\) \]}^{-\frac{\alpha}{2}} + \hat{I}_1(t)   \) \r\}  \right)},\; P_{\text{max}}\) \label{Eq:OptP1}\\
P_0^*(t) &= \min\( \frac{P_{c}}{W\(  P_{c} \exp\l\{ c_{1} -1 - \log\(\left(2v\sqrt{\hat{t}}\right)^{\alpha}\left[\sigma^{2}N^{-1} + \hat{I_{0}}(t)\right]\)   \r\}   \)}, \; P_{\text{max}}\) \label{Eq:OptP0} 
\end{align}
\small where $W(y)$ is a Lambert W function providing the solution $x$ of $y=x e^x$, which monotonically increases for $y>0$.

\hrulefill
\begin{align}
\hat{t}^* &= \frac{1}{4 v^2} \max\( \frac{1}{ \lambda_b  \(\beta^{-1}-1 \)}  ,\;   \[  \frac{P_0^*(t)}{\sigma^2 N^{-1} + \hat{I}_0(t)}\(\frac{P_1^*(t)}{ \sigma^2 N^{-1} {\lambda_b}^{-\frac{\alpha}{2}} + \hat{I}_1(t) }\)^{ \(1 + \frac{P_c}{P_1^*(t)}\)^{-1}-1 } \exp\l\{-c_1\(1 + \frac{P_c}{P_1^*(t)}\)^{-1}  \r\} \]^{\frac{2}{\alpha}} \)\label{Eq:OptHO}
\end{align}\normalsize
\end{figure*}

\subsection{EE under MF Interference}

The purpose of this subsection is to provide the closed-form $\EE_0(t)$ and $\EE_1(t)$. Deriving such closed-form expressions is viable with the aid of MF interference, i.e. interference normalized by BS density and the number of antennas \cite{UdnMFGSG:GC16}. Achieving this requires to show MF interference asymptotically converges to zero. 

The MF interference $\hat{I}_1(t)$ for $\EE_1(t)$ is not affected by BM mobility, and its convergence is proved by Theorem 1 in \cite{UdnMFGSG:GC16} under the following necessary conditions.
\begin{description}
\item \hspace{-35pt}{\textbf{A1}}. $\eta ,\; |\mu| <\infty$
\item \hspace{-35pt}{\textbf{A2}}. For $R<\infty$, $N {\lambda_b}^\alpha/{\lambda_u }^4 \rightarrow \infty$ and $\lambda_u \rightarrow \infty$
\item \hspace{-35pt}{\textbf{A3}}. For $R\rightarrow\infty$, $N {\lambda_b}^\alpha/\({\lambda_u R}\)^4 \rightarrow \infty$
\end{description}

The MF interference $\hat{I}_0(t)$ for $\EE_0(t)$ affected by BM mobility requires additional necessary conditions as follows.
\begin{proposition} \emph{(MF Interference under BM)} In a downlink cellular network with BM mobility following users, its MF interference $\hat{I}_0(t)$ converges under the fading condition \emph{\textbf{A1}} while satisfying either one of the following conditions.
\begin{description}
\item \hspace{-35pt}\emph{\textbf{A4}}. For $R<\infty$, $N /(\hat{t}v^2\lambda_u)^4 \rightarrow \infty$ and $\lambda_u \rightarrow \infty$
\item \hspace{-35pt}\emph{\textbf{A5}}. For $R\rightarrow\infty$, $N /\(\hat{t}v^2\lambda_u R\)^4 \rightarrow \infty$
\end{description}
\end{proposition}
It is worth noticing that \textbf{A4} and \textbf{A5} include $\hat{t}$ instead of $t$ since the HO interval is the maximum moving time span. In addition, $N /(\hat{t}v^2\lambda_u)^4 \rightarrow \infty$ can be interpreted as: (i) massive MIMO where $N\rightarrow \infty$, (ii) freeway vehicular networks where $\hat{t}\rightarrow 0$, and/or (iii) user stationary UDNs where $v\rightarrow 0$. For simplicity, we confine our interest to (i) under the sectorized antenna model as specified in Section II. Incorporating a realistic massive MIMO channel modeling and/or other interpretations could lead to another interesting topic for the further extensions.

Utilizing $\hat{I}_0$ and $\hat{I}_1$ provides closed-form EE as follows.
\begin{proposition} \emph{(EE with and without HO)} In a downlink cellular network with users having BM mobility while ensuring \emph{\textbf{A1}--\textbf{A5}}, a typical user's EE at $t>0$ with HO $\EE_1(t)$ and without HO $\EE_0(t)$ are respectively given \eqref{Eq:EE1} and \eqref{Eq:EE0} at the bottom of this page.
\end{proposition}
As described in Section III-B, $\EE_1(t)$ contains $\exp\(-\lambda_b A(\theta,t)/N_p\)$ that captures pilot response congestion that increases the association distance. When the response from the nearest BS collides with others, the user resorts to associating with the next nearest BS, degrading $\EE_1(t)$. 

Decreasing pilot broadcast area  $A(\theta,t)$ by increasing the pilot RSRP threshold $\theta$ mitigates this problem, but in return increases the occurrence of no response from BSs. The latter affects retransmissions that consumes $\varepsilon$ time. Such a trade-off is captured by Fig. 3 in Section V.

\subsection{User-Centric Mobility Management}
By utilizing the closed-form $\EE_0(t)$ and $\EE_1(t)$, this subsection derives the EE-optimal transmit powers $P_0^*(t)$ and $P_1^*(t)$ as well as the optimal HO time window $\hat{t}^*$ in tractable forms.

\begin{proposition} \emph{(Optimal Transmit Power)} Under the conditions \textbf{A1}--\textbf{A5}, $\EE_1(t)$ maximizing transmit power is given as \eqref{Eq:OptP1}, and $\EE_0(t)$ maximizing power is given as \eqref{Eq:OptP0} at the bottom of the next page.
\end{proposition}
Note that the above transmit powers contain recursive calculations since $\hat{P}_i(t)$ interacts with $\hat{P}_i(t)$ in \eqref{Eq:I_i}. Its quick convergence to a unique $\hat{P}^*(t)$ within a couple of iterations is verified in \cite{UdnMFGSG:GC16}.

Combining Propositions 2 and 3 yields the following optimal HO time window.
\begin{proposition} \emph{(Optimal HO Time Window)} Under the conditions \textbf{A1}--\textbf{A5}, at a typical user, $\EE_1^*(t)>\EE_0^*(t)$ satisfying optimal HO time window $\hat{t}^*$ is given as \eqref{Eq:OptHO} at the bottom of this page.
\end{proposition}
Note that $\hat{t}^*$ is lower bounded by $1/\(4v^2 \lambda_b \[\beta^{-1}-1\] \)$ to avoid wasteful HO trials as described in Section III-B.

\begin{figure}\centering
\includegraphics[width=7cm]{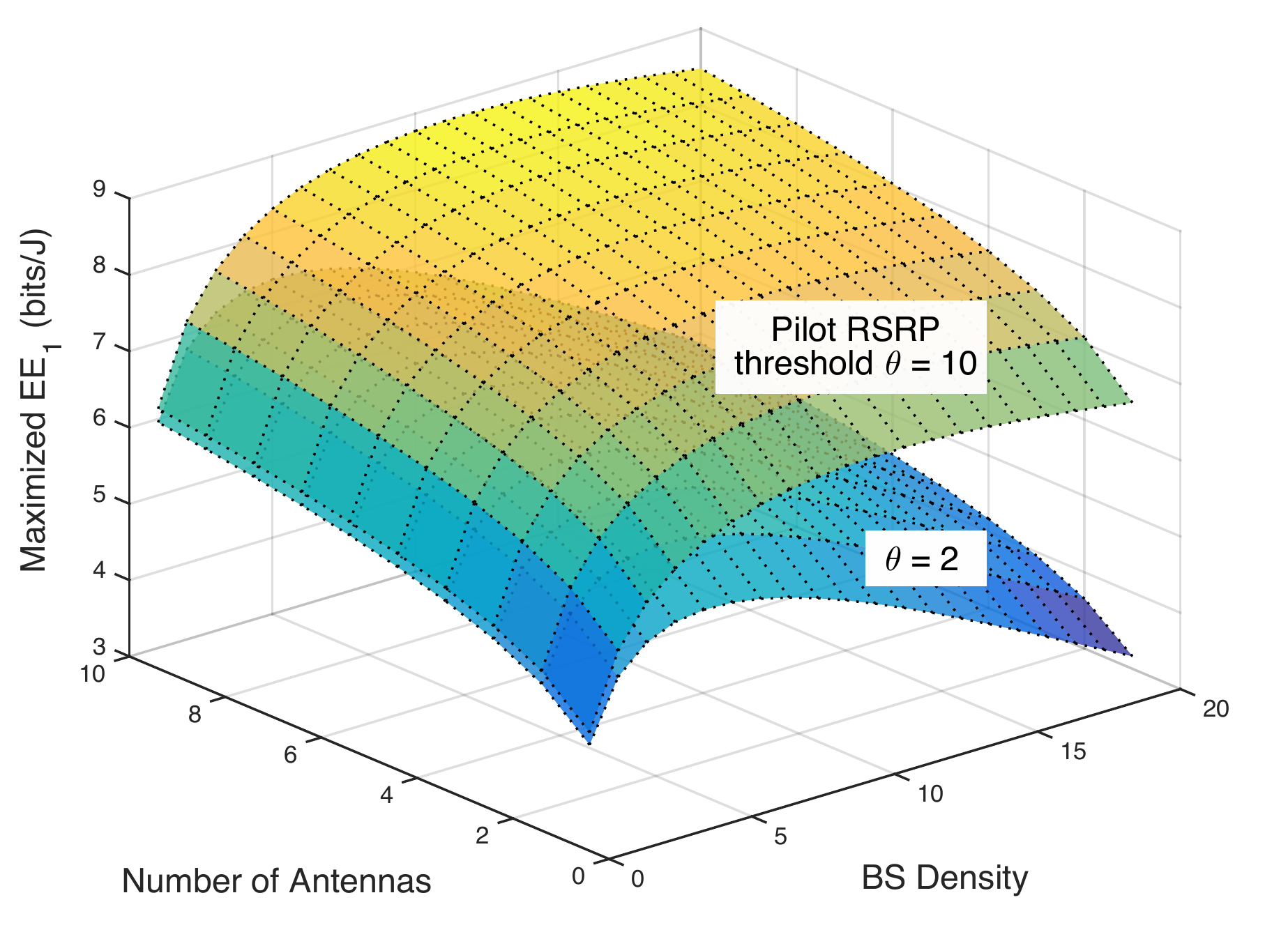}
	\caption{Maximized $\EE_1(t)$ w.r.t. the number of antennas and BS density.}
\end{figure}

\begin{figure}\centering
\includegraphics[width=7cm]{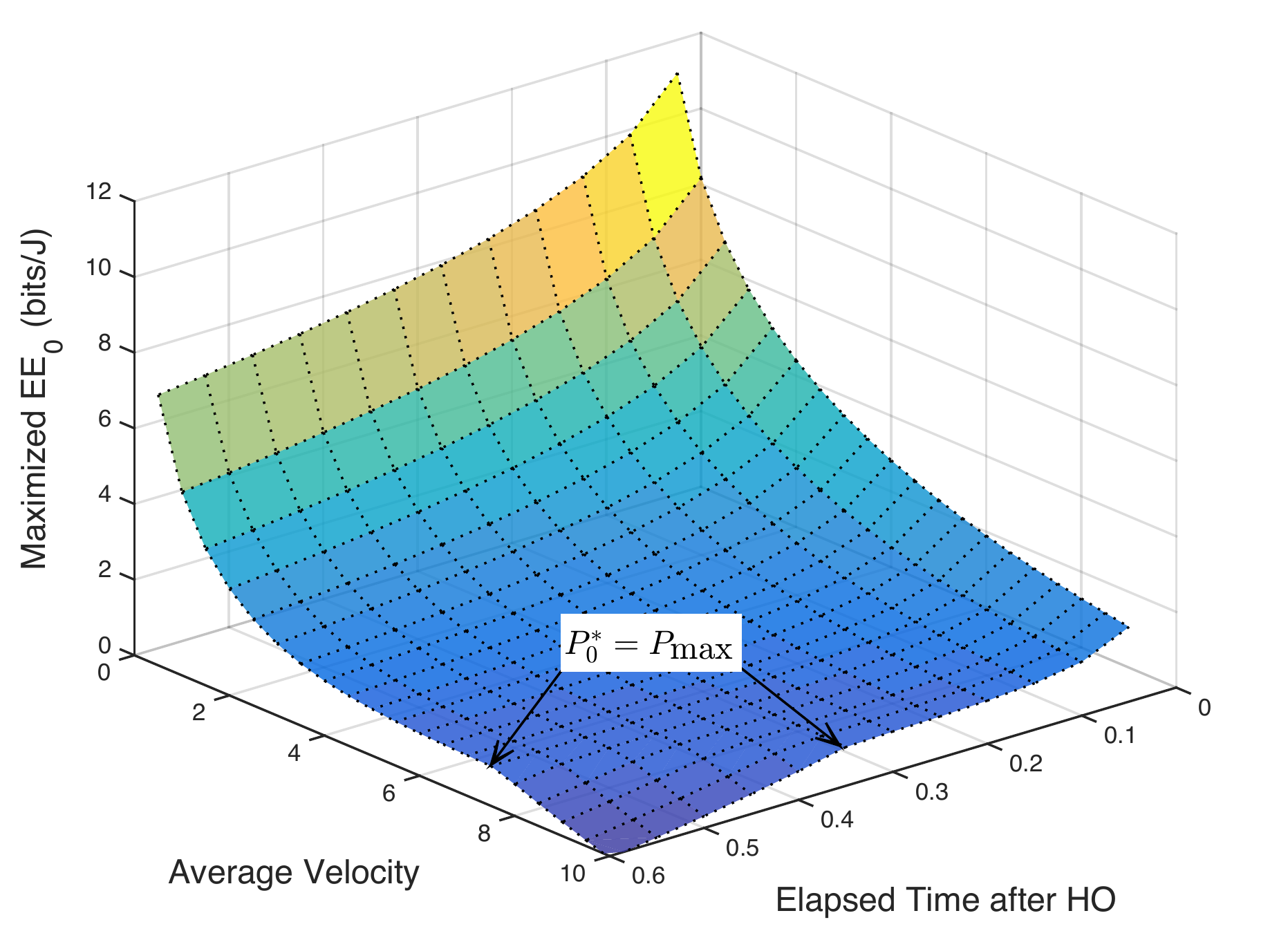}
	\caption{Maximized $\EE_0(t)$ w.r.t. elapsed time after HO and average velocity.}
\end{figure}

\section{Numerical Evaluation}
\small System parameters are given as follows: $\eta=|\mu|=0.01$, $\sigma^2=1$, $\lambda_b=20$, $P_{\text{max}}=20$, $N=10$, $\alpha=4$, $R=10$, $P_c=1$.\normalsize

Fig. 3 illustrates that $\EE_1^*(t)$ monotonically increases with $N$. The decreasing behavior for large $\lambda_b$ comes from the pilot response congestion that can be mitigated by increasing the pilot RSRP threshold $\theta$.

Fig. 4 indicates that $\EE_0^*(t)$ decreases along with the average velocity of users and the elapsed time after HO. This shows that fast users are prone to be handed over as the difference between $\EE_1^*(t)$ and $\EE_0^*(t)$ increases. For large average velocity and/or elapsed time, the optimal transmit power becomes $P_{\text{max}}$ so as to cope with the long association distance.

Fig. 5 shows that that the optimal HO time window $\hat{t}^*$ decreases with average velocity so that it can provide more frequent HOs. Moreover, the optimal HO time window is ensured to exceed the minimum HO time window (dotted red).

Fig. 6 verifies that the proposed mobility management provides $1.2$ times higher long-term average EE at a typical active BS given as $\EE_a^*(t)$, compared to a baseline model that has a fixed HO interval and transmit power. The figure also captures the impact of time-varying fading (thin) compared to its stationary fading (thick). The temporal channel dynamics with the given conditions slightly increase the optimal EE as well as the optimal HO time window. Its impact vanishes as time elapses, depicted by no difference in long-term average.

\section{Conclusion}
In this paper a decentralized user-centric association and power control policy were proposed. Spatio-temporal network dynamics were tractably incorporated in the proposed scheme by using MFG and SG. Based on the proposed mobility management in a downlink UDN, considering an uplink association rule could be an interesting topic.

\section*{Ackhowledgement}\small
This work was supported by the National Research Foundation of Korea (NRF-2014R1A2A1A11053234).
\pagebreak

\small
\section*{Appendix -- Proofs of Propositions}

\begin{figure}\centering
\includegraphics[width=7cm]{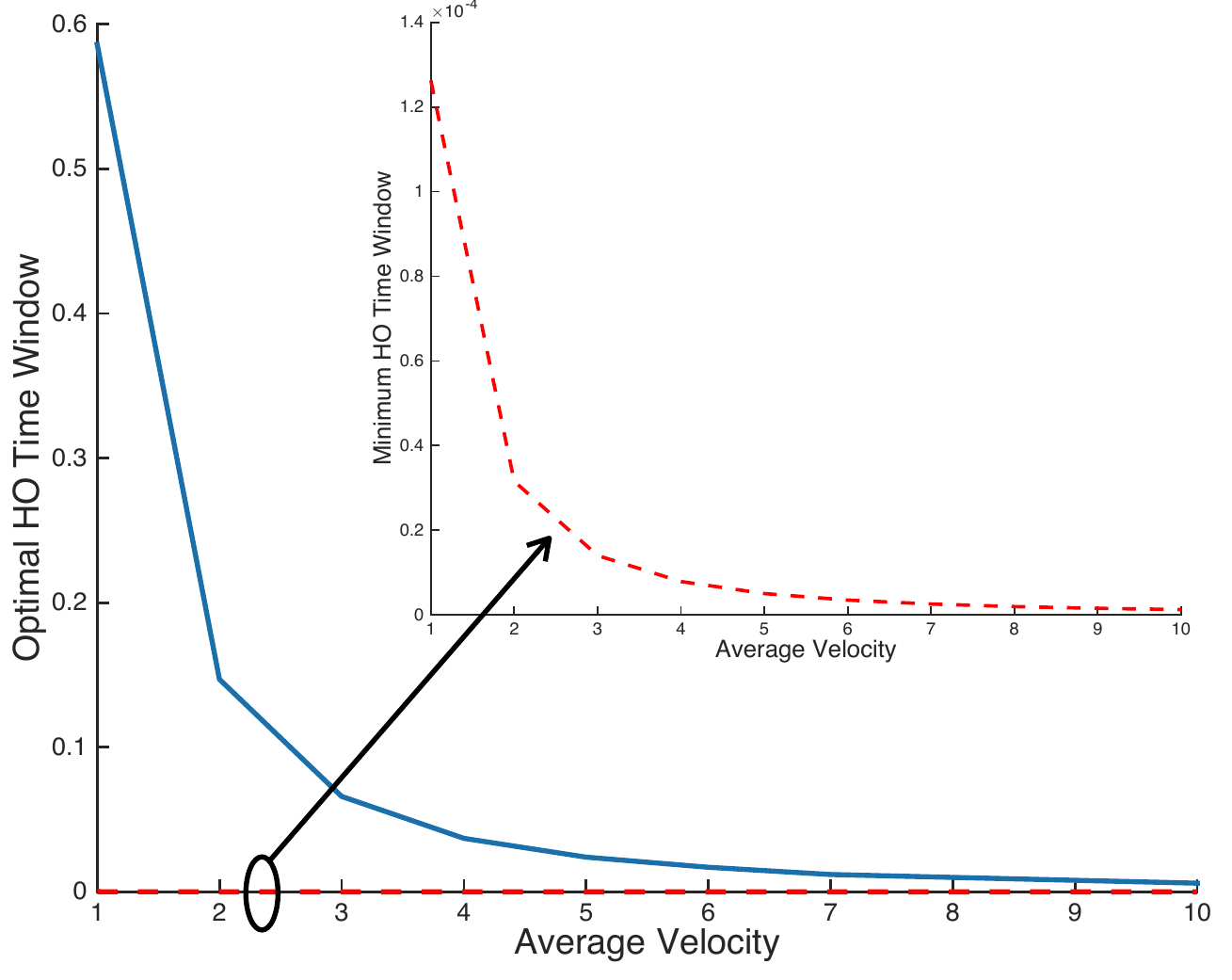}
	\caption{Optimal HO time window with respect to average velocity.}
\end{figure}

\begin{figure}\centering
\includegraphics[width=7.6cm]{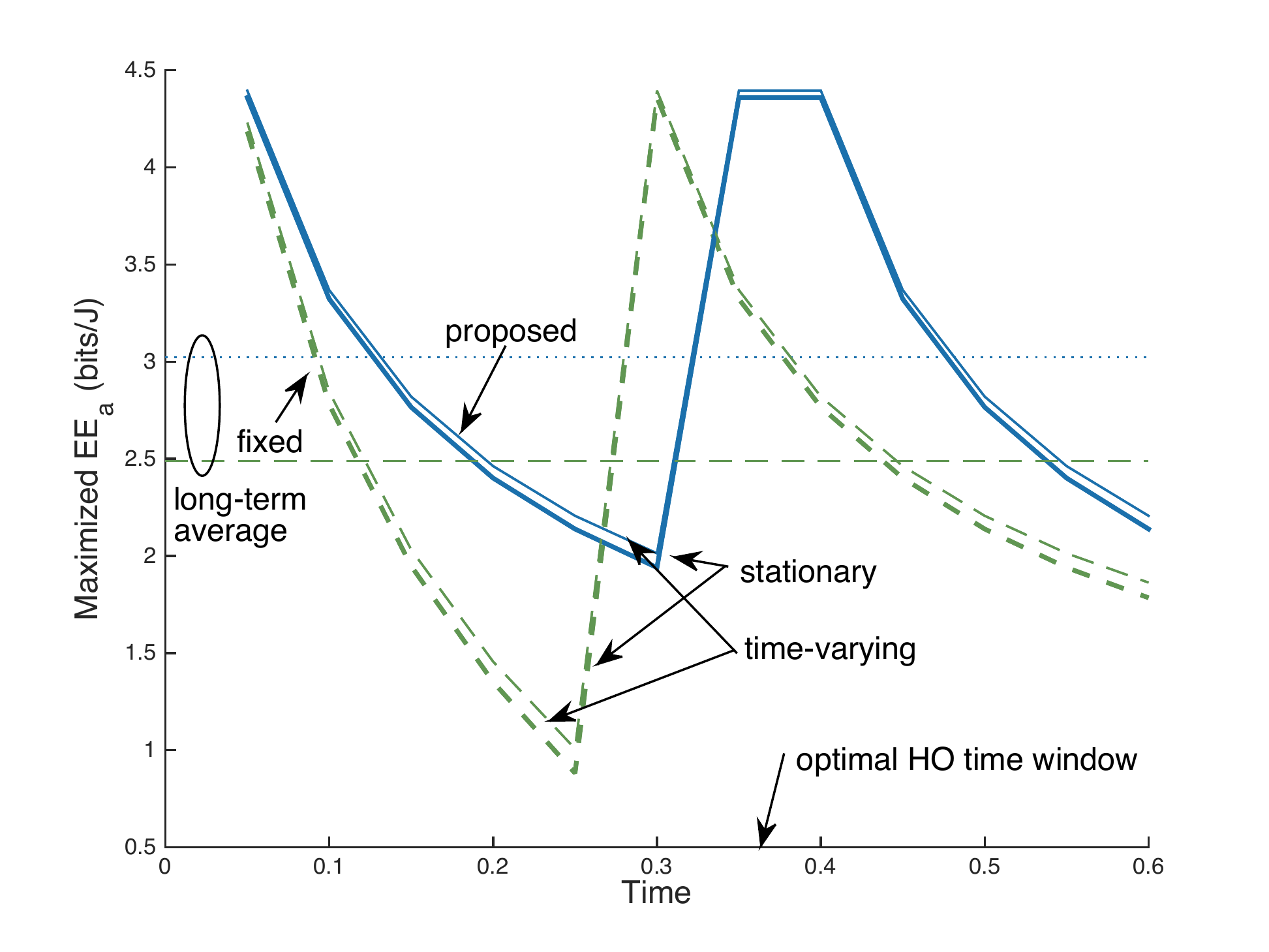}
	\caption{Maximized $\EE_a(t)$ with/without the proposed mobility management ($v=3$, $|\mu|=\sqrt{0.5}$, $\eta=0.5$; for the baseline, $\hat{t}=0.3$, $P_0(t)=0.25$).}
\end{figure}

\subsection{Proof of Proposition 1}
Let $\SINR_0(t)$ denote the signal-to-interference-plus-noise ratio when keeping the previous association, given as

\vspace{-10pt}\small\begin{align}
\hspace{-10pt}\SINR_0(t) &\overset{(a)}{=} N \cdot S\(t, \frac{1}{4 \hat{t} v^2}\)\[\sigma^2 + \frac{\theta_N}{2\pi}N\cdot I(t, \lambda_u) \]^{-1} \\
&=  S(t, 1)\bigg[ \frac{\sigma^2 \(4 \hat{t} v^2\)^\frac{\alpha}{2}}{N} + \underbrace{\frac{\( 4 t v^2 \lambda_u\)^{\frac{\alpha}{2}}}{\sqrt{N}}\cdot I(t, 1)}_{\hat{I}_0(t)} \bigg]^{-1} \label{Eq:SINRbm} 
\end{align}
\vspace{-10pt}

\noindent where $S(t, \lambda)$ and $I(t, \lambda)$ respectively denote the received power and interference at a typical user under the BS density $\lambda$ with a single BS antenna. The last step comes from the mapping theorem \cite{HaenggiSG}. For the step $(a)$, the locations of interferers providing HO at $t$ follow a homogeneous PPP thanks to Lemma 1, and the locations of interferers not providing HO also follow a homogeneous PPP by inverting Lemma 1 for BSs. The superposition of these homogenous PPPs yields another homogeneous PPP with density $\lambda_u$, leading to the result. Exploiting Theorem 1 in \cite{UdnMFGSG:GC16} and applying \eqref{Eq:SINRbm} yields $\hat{I}_0(t)$ that converges to $0$. \endproof

\subsection{Proof of Proposition 2}
Let $\SINR_1(t)$ denote the signal-to-interference-plus-noise ratio at the time instant when HO occurs, given as

\vspace{-10pt}\small\begin{align}
\SINR_1(t) &=  S(t, 1) \bigg[ \frac{\sigma^2}{{\lambda_b}^{\frac{\alpha}{2}}N} + \underbrace{\frac{1}{\sqrt{N}}\(\frac{\lambda_u}{\lambda_b}\)^{\frac{\alpha}{2}} I(t, 1)}_{\hat{I}_1 (t)}\bigg]^{-1}. \label{Eq:SINR}
\end{align} 
\vspace{-10pt}

Applying Theorem 1 in \cite{UdnMFGSG:GC16} proves the convergence of $\hat{I}_1(t)$ to $0$. Plugging $\hat{I}_1(t)$ and $\hat{I}_0(t)$ with applying Lemma 1 and Proposition 1 in \cite{UdnMFGSG:GC16} leads to the desired results. For the pilot response congestion, a single response success probability is $\exp\(-\lambda_b A(\theta,t)/N_p\)$ when uniformly selecting pilot resource block out of $N_p$. This success probability makes the number of retransmissions follow a geometric distribution, finalizing the proof.\endproof


\subsection{Proof of Proposition 3}

The local optimum of the objective function \eqref{Eq:EE1} satisfies Karush-Kuhn-Tucker (KKT) necessary conditions. Applying the first-order necessary condition yields the local optimal power $P_1^*(t)$ given as \eqref{Eq:OptP1}. Consider the second derivative test for \eqref{Eq:EE1}. Since $\ddot{\EE}_{1}(t)$ is monotonically increasing on the interval $[0,P_{\text{max}}]$ and $\ddot{\EE}_{1}(t)_{\vert{P_{1}(t)=P_1^*(t)}}<0$, the objective function monotonically increases on the interval $[0, \min{(P_1^*(t),P_{\text{max}})}]$ and monotonically decreasing on the interval $[\min{(P_1^*(t),P_{\text{max}})}, \max{(P_1^*(t),P_{\text{max}})}]$. This validates the derived local optimum is the global optimum. The procedure for $P_0^*(t)$ is identical to this, completing the proof. \endproof

\subsection{Proof of Proposition 4}
According to Lemma 2 and the distance to the nearest BS \cite{HaenggiSG} provides the minimum HO time window. Applying the results of Proposition 3 to $\EE_1(t)=\EE_0(t)$ provides $\hat{t}^*$. \endproof

\bibliographystyle{ieeetr}

\end{document}